\documentclass[10pt]{iopart}

\usepackage{graphicx}  % needed for figures
\usepackage{bm}        % for math
\usepackage[english]{babel}
\usepackage{amsfonts,amssymb,mathrsfs}
\usepackage{epstopdf}

\newcommand{\eq}{\begin{equation}}
\newcommand{\eeq}{\end{equation}}
\newcommand{\be}{\begin{equation}}
\newcommand{\ee}{\end{equation}}
\newcommand{\bea}{\begin{eqnarray}}
\newcommand{\nn}{\nonumber}
\newcommand{\eea}{\end{eqnarray}}

\begin{document}

\title{A no-go theorem for polytropic spheres in Palatini $f(R)$ gravity}

\author{Enrico Barausse${}^1$, Thomas P.~Sotiriou${}^1$ and John C.~Miller${}^{1,2}$}

\address{${}^1$SISSA, International School for Advanced Studies, 
Via Beirut 2-4, 34014 Trieste, Italy and INFN, Sezione di Trieste}\address{${}^2$Department of Physics (Astrophysics), University of Oxford, Keble
Road, Oxford OX1 3RH, England}
\eads{\mailto{barausse@sissa.it},\mailto{sotiriou@sissa.it},\mailto{miller@sissa.it}}
\date{\today}
\begin{abstract}
 Non-vacuum static spherically-symmetric solutions in Palatini $f(R)$ gravity
are examined. It is shown that for generic choices of $f(R)$, there 
are commonly-used equations of state for which no
satisfactory physical solution of the field equations can be found
within this framework, apart from in the special case of General
Relativity, casting doubt on whether
Palatini $f(R)$ gravity can be considered as giving viable alternatives to General Relativity.
 \end{abstract}

%\pacs{04.80.Cc, 04.20.Jb, 04.40.Dg}
%\maketitle

The search for theories of gravity which can serve as an alternative to General 
Relativity (GR) has received a powerful stimulus from current developments in 
observational cosmology. If one accepts GR as the correct theory of gravity, 
then it would seem that the energy density of the universe must be dominated at 
the present time by a cosmological constant, or by an unknown form of energy 
(dark energy) that mimics the behaviour of a cosmological constant~\cite{obs}. 
The various problems associated with this~\cite{carrol_review} have prompted many 
authors to question whether GR is indeed a completely correct theory of gravity 
on the classical level and to investigate possible alternatives which would not 
require the inclusion of dark energy.

One of these alternatives is $f(R)$ gravity in the Palatini formalism,
or simply Palatini $f(R)$ gravity~\cite{buch}, and we will be
examining some of its properties in this paper. We
first recall how this class of theories comes about as a
generalization of GR. The Einstein equations can be derived from
the Einstein--Hilbert action not only by the standard metric
variation but also by making independent variations with respect to
the metric and the connections (see, for example, \cite{wald}). This
is called Palatini variation and, if one proceeds in this way, the
form of the connections is a derived property rather than being
specified separately ({\it e.g.} by specifying the Levi-Civita
connection, as in standard GR). In Palatini $f(R)$ gravity, this
extended form of variation is applied to an action which is 
a general function of the scalar curvature $f(R)$:
 \be
\label{action}
S=\frac{1}{16\,\pi}\int d^4 x\sqrt{-g}f(R)+ S_M(g_{\mu\nu},\psi),
\ee
 where $R=g^{\mu\nu}R_{\mu\nu}$, $g$ is the determinant of the metric
$g_{\mu\nu}$, $S_M$ is the matter action and $\psi$ collectively
denotes the matter fields. (We are using, throughout, units in which
$c = G = 1$.)  Note that here $R_{\mu\nu}$ is not the Ricci tensor of
the metric $g_{\mu\nu}$, but is constructed from the independent
connections $\Gamma^{\lambda}_{\phantom{a}\mu\nu}$.

Independent variation with respect to the metric and the connections 
gives
 \bea
\label{field1}
F(R) R_{\mu\nu}-\frac{1}{2}f(R)g_{\mu\nu}&=&8\,\pi \, T_{\mu\nu},\\
\label{field2}
\nabla_\sigma(\sqrt{-g} F(R)g^{\mu\nu})&=&0,
\eea
 where $F(R)=\partial f/\partial R$, $T_{\mu\nu}\equiv
-2(-g)^{-1/2}\delta S_{M}/\delta g^{\mu\nu}$ is the usual
stress-energy tensor of the matter and $\nabla_\mu$ is the covariant
derivative with respect to the connections
$\Gamma^{\lambda}_{\phantom{a}\mu\nu}$. Setting $f(R)=R$ leads to the
standard GR Einstein equations while \eref{field2} in this case
incorporates the definition of the Levi-Civita connection. Note that
in order to derive \eref{field1} and~\eref{field2}, one assumes
that the matter action does not depend on the independent connections
[see \eref{action}]. Using Palatini variation while
allowing the matter to couple to the independent connections leads to
metric-affine $f(R)$ gravity~\cite{sotlib} which is a different theory
with enriched phenomenology~\cite{sot1, sot2}.

Since it was shown that Palatini $f(R)$ gravity might be able to address 
dark-energy problems~\cite{vollick}, a number of studies have been made of its 
cosmological aspects and of its consistency with cosmological 
constraints~\cite{palcosm} as well as of the Newtonian and post-Newtonian 
limits and consistency with Solar System constraints~\cite{newt}. However, 
the question of obtaining solutions describing stars and compact objects has 
received only a small amount of attention so far. In this paper, we focus on 
the problem of finding consistent solutions for static spherically-symmetric 
stars when $f(R)\ne R$.

We first note that doing this is helped by the fact that Palatini $f(R)$ 
gravity retains a useful characteristic of GR: the exterior spherically 
symmetric solution is unique (\textit{Birkhoff's theorem}). 
To see this one must 
take the trace of \eref{field1},
 \begin{equation}
\label{trace}
F(R)R - 2f(R) = 8\,\pi\, T,
\end{equation}
 where $T\equiv g^{\mu\nu}T_{\mu\nu}$. For a given $f(R)$, this is an
algebraic equation in $R$ and therefore it can be solved to give $R$
as a function of $T$. We will not consider cases where this equation
has no roots, since those do not give viable classical gravity
theories~\cite{franc}. Equation~\eref{trace} also implies that if $T=0$,
$R$ must be constant. If we denote the value of $R$ when $T=0$ by
$R_0$ and insert it into \eref{field2}, this equation reduces to
the covariant conservation of $g^{\mu\nu}$, fixing the %independent
connection to be the Levi-Civita one. This will be the situation in
vacuum and in this case \eref{field1} reduces to
 \be\label{eq:desitter}
\widetilde{R}_{\mu\nu}-\Lambda(R_0)g_{\mu\nu}=0,
\ee
 where $\widetilde{R}_{\mu\nu}$ is now indeed the Ricci tensor of the metric 
and $\Lambda(R_0)=R_0/4$. According to whether $R_0$ is zero or not, which of 
course depends on the choice of $f(R)$, the theory reduces in vacuum to GR 
without or with a cosmological constant. The vacuum spherically symmetric 
solution is unique in either case, being either Schwarzschild or 
Schwarzschild-(anti-)de Sitter.

Having determined an exterior solution, we then need to find an
interior solution and perform a matching between the two. Recently,
the generalisation of the Tolman--Oppenheimer--Volkoff (TOV)
hydrostatic equilibrium equation for Palatini $f(R)$ gravity has been
derived~\cite{TOV}, opening the way for finding equilibrium interior
solutions. We will consider here the matching of such interior
solutions with exterior ones.

We begin by reviewing the formulae that we will need. Denoting the
Ricci scalar of $g_{\mu\nu}$ by $\widetilde{R}\equiv
g^{\mu\nu}\widetilde{R}_{\mu\nu}$ and setting
$\widetilde{G}_{\mu\nu}=\widetilde{R}_{\mu\nu}-
g_{\mu\nu}\widetilde{R}/2$, the field equations~\eref{field1}
and~\eref{field2} can be rewritten as a single one
 \bea
\label{eq:field}
\widetilde{G}_{\mu \nu} \!&=&\! \frac{8\pi}{F}T_{\mu \nu}- \frac{1}{2}g_{\mu \nu}\! 
                       \left(\!R - \frac{f}{F} \right)\! +\! \frac{1}{F} \left(
			\widetilde{\nabla}_{\mu} \widetilde{\nabla}_{\nu}
			\!- g_{\mu \nu} \widetilde{\Box}
		\right)\! F-\nn\\
&& \quad- \frac{3}{2}\frac{1}{F^2} \left(
			(\widetilde{\nabla}_{\mu}F)(\widetilde{\nabla}_{\nu}F)
			- \frac{1}{2}g_{\mu \nu} (\widetilde{\nabla}F)^2
		\right),
\eea
 where $\widetilde{\nabla}_{\mu}$ is the covariant derivative with
respect to the Levi-Civita connection of $g_{\mu\nu}$ and
$\widetilde{\Box}\equiv
g^{\mu\nu}\widetilde{\nabla}_{\mu}\widetilde{\nabla}_{\nu}$.  To
arrive at this equation, one has to solve \eref{field2} for
$\Gamma^{\lambda}_{\phantom{a}\mu\nu}$, insert this into
\eref{field1} and express the resulting equation 
in terms only of metric quantities (for an alternative 
method, see \cite{sot1}).

Using the static spherically symmetric ansatz
\begin{equation}
\label{metric}
	ds^2 \equiv -e^{A(r)}{\rm d}t^2 + e^{B(r)}{\rm d}r^2 + r^2{\rm d}\Omega^2,
\end{equation}
 in \eref{eq:field}, considering perfect-fluid matter with
$T_{\mu\nu}=(\rho+p)u_\mu u_{\nu}+pg_{\mu\nu}$ (where $\rho$ is the
energy density, $p$ is the pressure and $u^\mu$ is the fluid
4-velocity) and representing $d/dr$ with a prime, one arrives at the
equations
 \bea\label{eq:Ap}
A' & = & \frac{-1}{1 + \gamma} \left(
	\frac{1 - e^B}{r} - \frac{e^B}{F}8\pi Grp
			+ \frac{\alpha}{r}
		\right), \\
	B' & = &  \frac{1}{1 + \gamma} \left(
			\frac{1 - e^B}{r} + \frac{e^B}{F}8\pi Gr\rho
			+ \frac{\alpha + \beta}{r}
		\right),\\
\alpha  &\equiv&  r^2 \left(
			\frac{3}{4}\left(\frac{F'}{F}\right)^2  + \frac{2F'}{rF}
			+ \frac{e^B}{2} \left( R - \frac{f}{F} \right)
		\right), \\
	\beta  &\equiv&  r^2 \left(
			\frac{F''}{F} - \frac{3}{2}\left(\frac{F'}{F}\right)^2
		\right),\qquad
\gamma \equiv \frac{rF'}{2F}.\label{eq:abc}
\eea
Defining $m_{\rm tot}(r)\equiv r(1 - e^{-B})/2$ and using Euler's 
equation,
%
%\begin{equation}\label{eq:euler}
%p'=-\frac{A'}{2}(p+\rho)
%\end{equation}
%
one gets the generalized TOV equations~\cite{TOV}:
\bea
\label{eq:OVB}
p' &=& -\frac{1}{1 + \gamma}\,\frac{(\rho + p)}{r(r - 2m_{\rm tot})}\left( m_{\rm tot} + \frac{4\pi r^3 p}{F}  
          - \frac{\alpha}{2} (r - 2m_{\rm tot} ) \right),\\
\label{eq:mass}
m_{\rm tot}' &=& \frac{1}{1 + \gamma} \bigg(\frac{4\pi r^2\rho}{F} + 
\frac{\alpha\!+\!\beta}{2}
- \frac{m_{\rm tot}}{r}(\alpha\! +\! \beta\! - \!\gamma) \bigg).
\eea
 We consider here matter which can be described by a one-parameter
equation of state (EOS) $p = p(\rho)$. When this is specified, one can
in principle solve the above equations and derive an interior solution. However,
this is hard to do in practice because the equations are implicit,
their right-hand sides effectively including through $F'$ and $F''$
both first and second derivatives of the pressure, \textit{e.g.}, $F'=
d/dr\,[F(R(T))]=(dF/dR)\,(dR/dT)\,(dT/dp)\,p'$. We therefore first put
them in an explicit form, which allows us not only to solve them
numerically, but also to study their behaviour at the stellar surface.

Multiplying \eref{eq:OVB} by $dF/dp$ and using the definitions of
$\alpha$ and $\gamma$, we get a quadratic equation in $F'$, whose
solution is
  \be
\label{eq:F1}
  F'=\frac{-4 r F ({\cal C}-F) (r-2 m_{\rm tot})+D\sqrt{2\Delta}}{r^2
(3 {\cal C}-4 F) (r-2 m_{\rm tot})}
  \ee
where $D=\pm1$ and where we have defined 
\bea 
{\cal{C}}&=&\frac{dF}{dp}(p+\rho)=\frac{dF}{d\rho}\frac{d\rho}{dp}(p+\rho),
\label{eq:calC}\\
\Delta&=&F r^2 (r-2 m_{\rm tot}) \left[8 F ({\cal C}-F)^2 
(r-2 m_{\rm tot})\right.-\\
& &-\left.{\cal C} (4 F-3
     {\cal C}) \left((16 \pi  p-F R+f) r^3+4 F m_{\rm tot}\right)\right].\nn
\eea
We will now focus on polytropic EOSs given by 
$p = \kappa {\rho_0}^{\Gamma}$, where $\rho_0$ is the rest-mass density and
$\kappa$ and $\Gamma$ are constants, noting that this can be rewritten
as $\rho=(p/\kappa)^{1/\Gamma}+p/(\Gamma-1)$, giving a direct link
between $p$ and $\rho$. In \eref{eq:calC}, we have written $\cal
C$ in terms of $dF/d\rho$ because this is finite at the stellar
surface ($r = r_{\rm out}$ where $p = \rho = 0$). In fact,
$dF/d\rho=(dF/dR)\,(dR/dT)\,(3dp/d\rho-1)$, where $dF/dR$ and $dR/dT$
are in general finite even when $T=3p-\rho$ goes to zero [check for
instance the $R^2$ or $1/R$ models] and $dp/d\rho\to0$ for $p\to0$. 
Note also that while $d\rho/dp$ diverges when $p\to0$, the product 
$(p+\rho)\,{d\rho}/{dp}$ goes to zero for $p\to0$ if $\Gamma<2$.
Therefore, for a polytrope with $\Gamma<2 $, ${\cal C}=0$ at the
surface. 

We now match the interior solution to the exterior one. For the
latter, the general solution to \eref{eq:desitter} is
$\exp(-B(r))=\ell\exp(A(r))=1-2m/r-R_0 r^2/12$, where $\ell$ and $m$ are
integration constants to be fixed by requiring continuity of the
metric coefficients across the surface and $R_0$ is
again the vacuum value of $R$. Using the definition of $m_{\rm 
tot}(r)$ this gives, in the exterior,
% \begin{equation} \label{eq:m_ext} 
$m_{\rm tot}(r) = m+r^3R_0/24\;.$ 
%\end{equation} 
Besides continuity of the metric, one has to impose some junction conditions for $A'$.
The exterior solution evaluated at the surface gives
% Besides continuity of the metric, the Israel junction conditions also
%require continuity of $A'$, since $\rho=0$ at the matching surface and
%no surface layer approach can be adopted. For the exterior, at the surface
%one has
 %
\be
\label{eq:requested_A_prime}
A'(r_{\rm out})=\frac{2 \left(r_{\rm out}^3 R_0-12 m\right)}{r_{\rm out} 
\left(R_0 r_{\rm out}^3-12 r_{\rm out}+24 m\right)}\:,
 \ee
whereas the value of $A'(r_{\rm out})$ for the interior solution can be calculated
with \eref{eq:Ap}. For this we need
$F'(r_{\rm out})$. Evaluating \eref{eq:F1} at the surface, where
${\cal C}=p=0$ and $R$, $F$ and $f$ take their constant vacuum values
$R_0$, $F_0$ and $f_0=F_0 R_0/2$, we get
% 
%\begin{equation} \label{eq:F1_surface}
$F'(r_{\rm out})=-{(1+\widetilde{D})F_0}/{r_{\rm out}}\;,$
%\end{equation}
 % 
 where $\widetilde{D}=D\,{\rm sign}(r_{\rm out}-2 m_{\rm
tot})$~\footnote{Unlike in GR, one cannot prove that $r_{\rm
out}>2 m_{\rm tot}$ from \eref{eq:OVB} because $p'$ is not
necessarily positive, although one may expect $r_{\rm out}>2 m_{\rm
tot}$ in sensible solutions.}. 
Choosing $\widetilde{D}=1$ implies $\gamma=-1$ at the surface 
[see \eref{eq:abc}] giving $A'\to\infty$ for $r\to 
r_{\rm out}^{-}$ [see \eref{eq:Ap}], whereas $A'$ 
keeps finite for $r\to r_{\rm out}^{+}$ 
[see \eref{eq:requested_A_prime}]. Because $\widetilde{G}_{\mu\nu}$ 
involves $A''$, this infinite discontinuity leads to the presence of 
Dirac deltas in the field equations. These Dirac deltas
cannot be cancelled by the derivatives of $F'$ on the right-hand side, 
because the discontinuity of
$F'$ is only a finite one, and one should therefore invoke 
an infinite surface density at $r=r_{\rm out}$. 
Since this is unreasonable, we focus only on $\widetilde{D}=-1$, for which 
$F'(r_{\rm out})= 0$ when $r\to r_{\rm out}^-$, 
making both $F'$ and $A'$ continuous across
the surface.

In order to study the behaviour of $m_{\rm tot}$ at the surface, we
need first to derive an explicit expression for $F''$. If we take the
derivative of \eref{eq:F1}, $F''$ appears on the left-hand side
and also on the right-hand side [through $m_{\rm tot}'$, calculated from 
\eref{eq:mass} 
and the definition of $\beta$, \eref{eq:abc}], giving a linear 
equation in $F''$. The solution to this,
 evaluated at the surface, is
\be
\label{eq:F''}
 F''(r_{\rm out})=\frac {\left(R_0 r_{\rm out}^3-8 m_{\rm tot}\right) 
{\cal C}'}{8 r_{\rm out} (r_{\rm out}-2 m_{\rm tot})}
\ee
 Evaluating $\alpha$, $\beta$ and $\gamma$ at the surface using $F'=0$
and $F''$ given by \eref{eq:F''}, and inserting 
into \eref{eq:mass} gives
\begin{equation}
\label{eq:mprime} 
 m_{\rm tot}'(r_{\rm out})=\frac{2 F_0 R_0 r_{\rm out}^2+\left(r_{\rm 
out}^3 R_0-8 m_{\rm tot}\right) {\cal C}'}{16 F_0}\;.
 \end{equation}
 For $1<\Gamma<3/2$, ${\cal C}'=d{\cal C}/dp\,p'\propto d{\cal
C}/dp\,(p+\rho)\to 0$ at the surface so that expression
\eref{eq:mprime} is finite and it even gives continuity of $m_{\rm tot}'$
across the surface. %[\textit{cf.} \eref{eq:m_ext}]. 
However, for $3/2<\Gamma<2$, ${\cal C}'\to\infty$ as the surface is
approached, provided that $dF/dR(R_0)\neq0$ and
$dR/dT(T_0)\neq0$ 
(note that these conditions are satisfied by
 generic forms of  $f(R)$, {\em i.e.~}whenever an $R^2$ term or a term
 inversely proportional to $R$ is present).
 While $m_{\rm tot}$ keeps finite [as can be shown using 
$p\sim(r_{\rm out}-r)^{\Gamma/(\Gamma-1)}$, which can be derived by integrating \eref{eq:OVB} near the surface], 
the divergence of $m_{\rm tot}'$
drives to infinity the Riemann tensor of the metric,
$\widetilde{R}_{\mu\nu\sigma\lambda}$, and curvature invariants, such
as $\widetilde{R}$ or
$\widetilde{R}^{\mu\nu\sigma\lambda}\widetilde{R}_{\mu\nu\sigma\lambda}$,
as can easily be checked~\footnote{This seems to have been missed in
\cite{barraco}.}. This singular behaviour would cause 
unphysical phenomena, such as infinite tidal forces 
which would destroy
anything present at the surface [\textit{cf.} the geodesic deviation equation].
We can then conclude that no physically relevant solution exists for
any polytropic EOS with $3/2<\Gamma<2$. 
Of course, polytropes give only simplified models for stars and
the EOS in the outer layers is critical for the
behaviour of $m'_{\rm tot}$ at the surface in the non-GR case. One
would like to use a more accurate EOS, but 
while this can give regular solutions in many
cases (\textit{e.g.} if $p\propto\rho_0$ near the surface), 
the existence of counter-examples is worrying for the viability
of the theory.

Setting aside the surface singularity, we next turn to the behaviour in the 
interior, focusing on models of neutron 
stars constructed using an analytical approximation to the FPS 
EOS~\cite{haensel}. Adding positive powers of $R$ to the Einstein-Hilbert 
action produces significant effects for compact stars while adding negative 
ones predominantly affects more diffuse stars. Generically, though, one would 
expect terms of both types to be added if there is a deviation away from GR. 
Since the $1/R$ term commonly used in cosmology would have a negligible effect 
in the interior of a neutron star, we used here $f(R)=R+\epsilon R^2$.
As can be seen from \eref{eq:field}, the metric will be
sensitive to derivatives of the matter fields, since $R$ is a function
of $T$~\footnote{The unusual behaviour of this class of theories has
been mentioned in a different context in \cite{olmo2}. However,
we disagree with the claims made there about the violation of the
equivalence principle, because they seem to be based on an ill-posed
identification of the metric whose geodesics should coincide with
free-fall trajectories.}. 

%\begin{center}
\begin{figure}[t]
\hspace{63pt}
\includegraphics[width=11.cm]{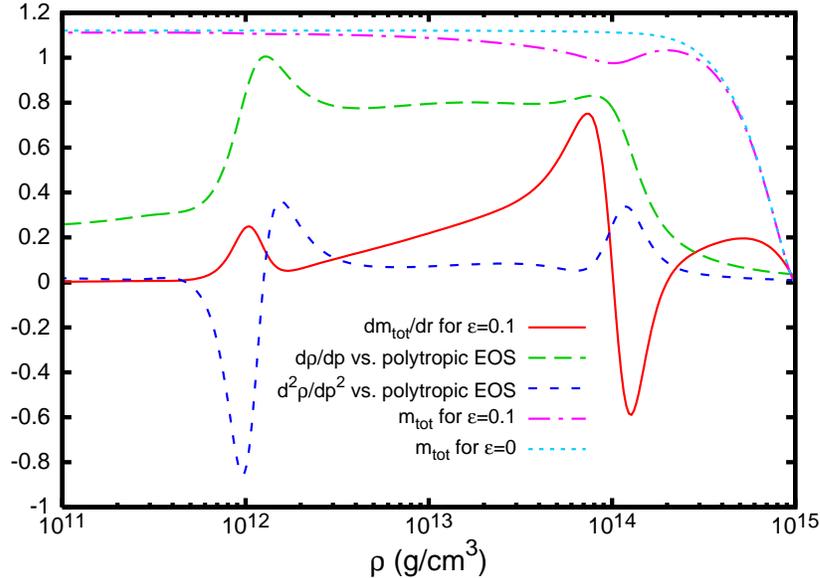}
\caption{\label{fig}Profiles of $m_{\rm tot}$ (in
$M_\odot$) and other associated quantities plotted against density
in the interior of a neutron-star with central density
$10^{15}{\rm g/cm^3}$ and $p'=0$ in the centre as required by local
flatness. We have used the FPS EOS~\cite{haensel}
and $f(R)=R+\epsilon R^2$. The dot-dashed
line shows $m_{\rm tot}$ as calculated with $\epsilon=0.1$ and
the dotted line shows the equivalent curve in GR
($\epsilon=0$); the solid line shows $dm_{\rm tot}/dr$
(in $M_\odot/{\rm km}$) for
$\epsilon=0.1$ (this value is orders of magnitude lower than Solar System constraints~\cite{thomas_constraints}). 
Note the bumps in the $dm_{\rm tot}/dr$ curve 
resulting from rapid composition changes in the EOS (the corresponding
features in the $m_{\rm tot}$ curve for $\epsilon=0.1$ are less
apparent but a noticeable dip is seen at $\rho\sim10^{14}{\rm
g/cm^3}$). To make evident the influence of composition changes, we
also show comparisons between the FPS EOS and a
polytrope (with $\Gamma=4/3$ and $\kappa=10^{15}$ cgs): the 
long-dashed and short-dashed curves show
$0.1\times(d\rho/dp)_{\rm FPS}/(d\rho/dp)_{\rm polytrope}$ and
$0.01\times(d^2\rho/dp^2)_{\rm FPS}/(d^2\rho/dp^2)_{\rm polytrope}$,
respectively.}
\end{figure}
%\end{center}
%
This can be seen in Figure~\ref{fig}: $m_{\rm
tot}$, which in GR has a smooth profile, now develops peculiar
features when $d\rho/dp$ and $d^2\rho/dp^2$ change rapidly in going
from the core to the inner crust and from the inner crust to the outer
crust. If $m_{\rm tot}$ were plotted against the radius, these
features would look much more abrupt, because they occur in a small
range of radii close to the surface. While $m_{\rm tot}$ does not 
represent a real mass in the interior, such a strong dependence of the
metric on the derivatives of the matter fields is not very plausible
and could have dramatic consequences. 

We have therefore found two unappealing characteristics of Palatini
$f(R)$ gravity as applied to stellar models, each of which 
arises because of the dependence of the metric on higher order
derivatives of the matter field. First: whether or not a regular
solution can be found depends crucially on the
microphysics, through the EOS, with polytropic EOSs having
$3/2<\Gamma<2$ being ruled out for generic $f(R)$. 
Second: even if an EOS does allow
a regular solution at the surface, the interior metric depends on the
first and second derivatives of the density with respect to the
pressure, giving a problematic behaviour. While polytropic EOSs are
highly idealized, we note that $\Gamma=5/3$, 
corresponding to an isentropic monatomic gas or a degenerate
non-relativistic particle gas,
falls within the range not giving a regular solution.
The fact that the gravity theory cannot provide a consistent 
description
for these cases, casts doubt on whether it should be considered  
as a viable alternative to GR.

Since the problems discussed here arise due to the dependence of the
metric on higher order derivatives of the matter fields, one can
expect that they will also appear in other gravity theories having
these characteristics. Any theory having a representation in which the
field equations include second derivatives of the metric and higher
than first derivatives of the matter fields will face similar problems
because having a higher differential order in the metric than in the
matter field is what guarantees that the metric depends in a
cumulative way on the matter. If this is not the case then the metric
loses its immunity to rapid changes in matter gradients since it is
directly related to them instead of being an integral over them.

 The same problem should be expected for any theory which includes
fields other than the metric for describing the gravitational
interaction (\textit{e.g.}~scalar fields) which are algebraically related
to matter rather than dynamically coupled. In this case one can
always solve the field equations for the extra field and insert the
solution into the field equation for the metric, inducing a dependence
of the metric on higher derivatives of the matter fields. An example of such a theory is a scalar-tensor theory with Brans-Dicke
parameter $\omega=-3/2$, which is anyway an equivalent representation
of Palatini $f(R)$ gravity~\cite{sot1}. One should mention that 
this problem could probably be addressed in Palatini $f(R)$ gravity by
adding higher order curvature invariants in the action [{\it
e.g.~}$f(R, R^{\mu\nu}R_{\mu\nu})$], since this would introduce more
dynamics and break the non-dynamical coupling between matter and the
extra gravitational degrees of freedom. 

In conclusion, we suggest that our results cast doubt on the viability of
theories including higher order derivatives of the matter fields in
one of their representations, such as Palatini $f(R)$ gravity
or $\omega=-3/2$ scalar-tensor theory.\\
\\
{\em Note added:}  Prior to publication, 
a paper by Kainulainen {\it et al.} appeared~\cite{finns}, which agreed with the validity of our results but
criticized our interpretation of them. A response to this criticism, as well as an additional analysis of the issues presented here, 
can be found in \cite{newpaper}.

\section*{Acknowledgements} We wish to thank G.~Olmo and
S.~Liberati for helpful discussions.

\end{document}